# Atomic cluster expansion force field based thermal property material design with density functional theory level accuracy in non-equilibrium molecular dynamics calculations over sub-million atoms

Takumi Araki[1*], Shinnosuke Hattori[1], Toshio Nishi[1], and Yoshihiro Kudo[1]

[1]Material Device Analysis Center, Third Research Department, Sony Semiconductor Solutions Corporation

*Corresponding author: Takumi.Araki@sony.com

**Abstract**

Non-equilibrium molecular dynamics (NEMD) techniques are widely used for investigating lattice thermal conductivity. Recently, machine learning force fields (MLFFs) have emerged as a promising approach to enhance the precision in NEMD simulations. This study is aimed at demonstrating the potential of MLFFs in realizing NEMD calculations for large-scale systems containing over 100,000 atoms with density functional theory (DFT)-level accuracy. Specifically, the atomic cluster expansion (ACE) force field is employed, using Si as an example. The ACE potential incorporates 4-body interactions and features a training dataset consisting of 1000 order structures from first-principles molecular dynamics calculations, resulting in a highly accurate vibrational spectrum. Moreover, the ACE potential can reproduce thermal conductivity values comparable with those derived from DFT calculations via the Boltzmann equation. To demonstrate the application of MLFFs to systems containing over 100,000 atoms, NEMD simulations are conducted on thin films ranging from 100 nm to 500 nm, with the 100 nm films exhibiting defect rates of up to 1.5%. The results show that the thermal conductivity deviates by less than 5% from DFT or theoretical results in both scenarios, which highlights the ability of the ACE potential in calculating the thermal conductivity on a large scale with DFT-level accuracy. The proposed approach is expected to promote the application of MLFFs in various fields and serve as a feasible alternative to virtual experiments. Furthermore, this work demonstrates the potential of MLFFs in enhancing the accuracy of NEMD simulations for investigating lattice thermal conductivity for systems with over 100,000 atoms.

# I. INTRODUCTION

Nanoscale thermal management is a crucial aspect for the development of thermoelectric devices [1–3]. Simulations of thermal transport have proven instrumental in achieving low thermal conductivity in thermoelectric materials [3–5]. Ab initio-based phonon calculations provide valuable insights into perfect bulk systems, which can enhance our understanding of this domain [6–8]. However, the presence of defects, grain boundaries, and thickness variations in realistic materials [9–12] render the simulations of large-scale systems challenging. For analyzing the thermal conductivity of materials at the nanoscale, factors such as the film thickness and presence of defects must be considered. Molecular dynamics (MD) has emerged as the preferred method for directly studying thermal conduction [13–16] in vertical space. Two key methods are available for calculating the thermal conductivity in MD simulations: nonequilibrium MD (NEMD) [11,13,15,16] and the Green–Kubo approach of equilibrium MD [14]. The accuracy of the force field in MD significantly influences the precision of thermal conductivity calculations [13,17,18]. Recent studies have achieved promising results using machine learning force fields (MLFFs) to accurately predict the thermal conductivity [19–23]. However, these calculations have been limited to systems with fewer than 100,000 atoms.

Different variants of MLFFs have been developed, such as the neural network potential [24,25], Gaussian approximation potential [23,26,27], moment tensor potential [28,29], and atomic cluster expansion (ACE) potential [30–33]. The ACE, well-known for its accuracy and computational speed, enables calculations involving more than 100,000 atoms [31]. ACE can take into account many-body effects to expand energy based on atomic cluster structure products. The ACE force field can be used to calculate thermal conductivity using the NEMD or Green–Kubo approach. Notably, the Green–Kubo approach is more computationally intensive than the NEMD [14,16,34], and the NEMD addresses the transfer of thermal energy through atomic vibrations for calculating the thermal conductivity.

Si has been widely used as a test material for predicting the thermal conductivity through simulations [11,12,14,15,23]. Various calculation methods, such as phonon calculations, equilibrium MD, and NEMD, have been noted to yield results consistent with experimental findings pertaining to changes in the thermal conductivity with the temperature and system size. Nevertheless, most of the existing studies have focused on systems with less than 100,000 atoms [15,23]. In this context, it is necessary to develop large-scale, detailed atomic computations to guide the enhancement of actual devices.

In this study, we performed MD simulations using an MLFF to validate the accuracy of the force field and its applicability to Si thin film and defect models. We focused on thermal conductivity calculations for systems with more than 100,000 atoms, emphasizing the achievement of high accuracy at the density functional theory (DFT) level. In the simulations, an MLFF constructed from ab initio MD (AIMD) for Si was used to verify the accuracy of vibrational behavior in phonon calculations and to evaluate the effects of the film thickness and vacancy rates in NEMD. Harmonic and anharmonic phonons were calculated in static phonon computations for interatomic forces associated with atomic displacements. The Si film thickness was varied from 100 nm to 500 nm in the NEMD simulations. Additionally, vacancy calculations for conducted with a fixed film thickness of 100 nm, with vacancy rates of up to 1.5%. In all cases, we compared the performance of the MLFF based on the ACE potential with those of ab initio calculations, experiments, and theoretical equations to verify the accuracy of harmonic and anharmonic phonons, along with NEMD results for three expansion orders: 2-body, 3-body, and 4-body. The objective was to demonstrate that a force field, formulated using a simple dataset such as AIMD with a set number of constituent atoms, retains sufficient information regarding atomic vibrations to accurately predict the thermal conductivity.

The remaining paper is structured as follows. Section II describes the dataset used to construct the ACE potential, phonon calculation conditions, and NEMD model. Section III describes the evaluations of the accuracy of the ACE potential for different expansion orders and number of atoms (Section III A), accuracy of harmonic and anharmonic phonons calculations (Section III B), and accuracy of thermal conductivity calculations of thin films and defects (Section III C). Section IV presents the concluding remarks.

## II. COMPUTATIONAL DETAILS

In the initial phase of this research, Si MLFFs were constructed based on DFT calculations using Quantum ESPRESSO V.6.7[35,36]. The DFT calculations were based on the Perdew–Burke–Ernzerhof exchange-correlation functional and Blochil's projector augmented-wave methods to generate the training data, with an energy cutoff of 50 Ry and $4 \times 4 \times 4$ k-mesh points. The atomic structure of Si consisted of 64 atoms, corresponding to a $2 \times 2 \times 2$ diamond lattice cell with lattice constants of 1.08 nm. In the AIMD calculations, the time step was set as 0.1 fs, and temperatures of 100 K, 300 K, and 500 K were considered. A total of 1000 data points were generated for each temperature, resulting in 3000 data points as the training dataset.

The ACE developed by Drautz [30] was used to construct the MLFF. ACE was chosen as it can achieve a better tradeoff between the accuracy and computational efficiency compared with other MLFFs [31]. In general, in ACE, the total energy is expressed as a polynomial expansion in atomic cluster configurations, such as pair, 3-body, 4-body, and so on. The ACE potential is defined by the order of the cluster configuration expansion (ν) and combination of quantum numbers $\{n,l,m\}$ in the basis, which represent the distributions of the clusters for radial and angular contributions [30]. The Si ACE potentials were constructed with expansion orders ranging from 2-body to 7-body (ν = 2, 3, …, 7), and their accuracies in terms of the energy and forces were compared. Detailed calculation conditions, such as the cutoff radius ($rc$) and quantum number $\{n,l,m\}$, for the final constructed potentials can be found in Table S1-1 in the Supporting Information. The Si ACE potentials were generated using PACE-maker [32] and integrated within the LAMMPS [37] MD simulation software package.

Atomic vibration spectral analysis and NEMD calculations were conducted in LAMMPS, a large-scale atomic/molecular massively parallel simulator [37]. In the NEMD calculation, we focused on a temperature of 300 K, given the abundant experimental and computational data available for this temperature [9,10,12,13,38–40]. For the vibrational spectral calculation, a bulk configuration containing 4096 Si atoms arranged in ideal 8×8×8 cells was used, with a time step of 0.1 fs, a calculation step of 100,000, and the canonical NVT ensemble. The vibration spectrum $G(v)$ was calculated as [41,42]

$$G(v) = \int_{-\infty}^{\infty} \frac{\sum_{i=1}^{N} \langle \mathbf{v}_i(t)|\mathbf{v}_i(0)\rangle}{\sum_{i=1}^{N} \langle \mathbf{v}_i(0)|\mathbf{v}_i(0)\rangle} e^{i2\pi vt} dt \qquad (1)$$

where $N$ represents the total number of atoms in the calculation system, and $v_i$ denotes the velocity of the $i$-th atom. The $G(v)$ of the system is proportional to the Fourier transform of the velocity autocorrelation function averaged over all atoms. This dynamical vibration spectrum was compared with the harmonic phonon density of states (PDOS) to confirm that the constructed ACE potential can capture phonon information. The NEMD simulation model included a thermal transport region with a variable film thickness $L$, heat source and sink regions sized approximately 5 nm each, and a fixed region of approximately 1 nm. $L$ corresponded to the Z direction, and the cross-sectional area of the model (approximately 4.32×4.32 nm$^2$) was aligned with the X-Y plane with periodic boundaries. Additional details of the model can be found in Figure S2-1 in the Supporting Information. A temperature gradient was generated by setting the heat source at 350 K and heat sink at 250 K, and a Langevin thermostat was used for calculating the thermal conductivity. Specifically, the thermal conductivity ($\kappa$) was calculated as [11,13,15,16]

$$\kappa = \frac{dQ}{dS} \times \frac{dL}{dT}, \qquad (2)$$

where $dQ$ is the heat flux computed by dividing the energy flow from the heat source and heat sink by the cross-sectional area ($dS$) and simulation time. The term $dL/dT$ represents the reciprocal of the temperature gradient. In the NEMD simulations, $L$ was varied from 100 nm to 500 nm, incorporating approximately 500,000 atoms (Table S2-1 in the Supporting Information presents the atomic counts of the model). The time step was 1 fs, equivalent time was 0.2 ns, non-equivalent time for the preliminary calculations was 0.3 ns, and span for thermal conductivity computations was 2 ns. The thermal conductivity was calculated every 1 ps, and the final value was obtained by averaging these values. The validity of the NEMD was evaluated by comparing the results with the Callaway–Holland model [39,40,43–45]. Additional information can be found in Section S3 in the Supporting Information.

Furthermore, phonon simulations were performed, and the results were compared with dynamical computations to validate the accuracy of the ACE potential. Harmonic and anharmonic phonons of Si were calculated using ALAMODE, a software application designed for analyzing lattice anharmonicity and thermal conductivity of solids [46]. In general, ALAMODE relies on external software, such as Quantum ESPRESSO and LAMMPS, to compute harmonic and anharmonic force constants. The phonon calculations were performed using a bulk configuration that contained 64 Si atoms with 2×2×2 diamond lattice cells. The base conditions for the DFT case, including the pseudopotential type, cutoff energy, and other variables, matched those used in the training session. ACE potentials with 2-body, 3-body, and 4-body expansion orders were used to estimate the harmonic and anharmonic force constants via LAMMPS. In the thermal conductivity calculations, the Boltzmann equation with a scattering relaxation time approximation in ALAMODE was applied [46–48]. The q-mesh points in the wavenumber space for the PDOS of the harmonic phonons and thermal conductivity associated with the anharmonic phonons were set as 40×40×40 points. We compared the PDOS to the dynamic vibration spectrum obtained from the MD calculations to ensure that the ACE potential possessed phonon information. Furthermore, the thermal conductivity obtained by the ACE potential with different expansion orders was compared with those derived using DFT calculations and experiments.

## Ⅲ. RESULTS AND DISCUSSION
### A. Effect of Si ACE potential expansion orders

The initial phase of our study involved the assessment of the accuracy of energy and interatomic forces associated with different expansion orders of the Si ACE potential. In the training process for the Si ACE potential, the weight of the interatomic forces was set as 0.99 in the loss function defined as in Equation 15 in Ref. [32]. This setting was aimed at focusing on the atomic vibration behavior and enhancing the reproducibility of interatomic forces. Figure 1a shows the root mean square error (RMSE) of the energy and interatomic forces, with the expansion order of the ACE potential varying from 2-body to 7-body. Detailed information regarding the expansion conditions and number of parameters can be found in Table S1-1 in the Supporting Information.

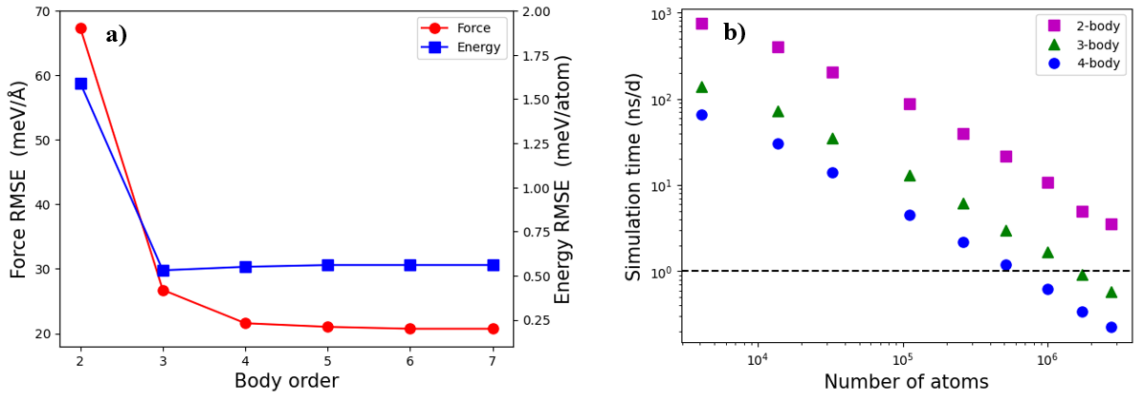

FIG. 1. (a) Root mean square error (RMSE) of force (red circles and line) and energy (blue squares

and line) derived from Si ACE potentials with different expansion orders. (b) Relationship between the number of atoms and simulation time per day. The dotted line shows a calculation of 1 ns per day.

The RMSE of the interatomic force begins converging at a body order of 4, yielding a force RMSE of 21.57 meV/Å. In contrast, the energy RMSE, estimated to be nearly 0.53 meV/atom, appears to converge from a 3-body order, as shown in Figure 1a. The increase in energy above the 4-body order indicates a larger weighting of the interatomic forces within the loss function. These findings suggest that a 3-body distribution in Si ACE potentials is adequate for representing energy. However, more complex interatomic forces necessitate an expansion order of 4 or more. Previous studies have demonstrated that a 4-body order is sufficiently accurate for constructing potentials for bulk materials such as Cu [32]. Thus, we concluded that a 4-body order is sufficient for describing the interatomic forces and compared the Si ACE potentials associated with 2-body to 4-body orders in terms of atomic vibrations. Notably, although our training results exhibit an energy RMSE accuracy that is an order of magnitude higher than that reported in Ref. [33] (RMSE=2.33 meV/atom), our potential is not superior. The increased accuracy is attributable to the larger amount of training data used, and the potentials in Ref. [33] demonstrate high transferability because of the incorporation of a large amount of Si data [49]. In comparison, our training set consisted of limited AIMD data, as our focus was on calculating the thermal conduction via atomic vibrations. In the application of MLFFs, the RMSE is not a reliable indicator. Instead, the quality of the force field must be assessed based on the physical property values derived from the calculations. Therefore, we examined the thermal conductivity resulting from the ACE potential. Subsequent analyses indicated that 3000 AIMD training data points are sufficient for ensuring the accuracy of thermal conductivity simulations.

Considering the abovementioned results, we constructed ACE potentials with expansion orders up to 4 and estimated the practical limit for the number of atoms in calculations, as illustrated in Figure 1b. Given that NEMD calculations for thermal conductivity necessitate approximately 1 ns of simulation time [11, 13], we evaluated the number of atoms that could be calculated at a rate of 1 ns per day. The computational resources used for this verification are presented in Section S4 in the Supporting Information. The 2-body and 3-body ACE potentials were noted to be computationally efficient, enabling a simulation involving over 1,000,000 atoms per day in 1 ns. In contrast, calculations involving 4-body potentials require more parameters and computational effort, and the optimal model size is approximately 500,000 atoms per day. Consequently, we used a 500,000-atom model as the central model for the NEMD calculations.

**B. Comparison of harmonic and anharmonic phonon calculations with DFT**

The objective of this analysis was to examine whether the Si ACE potential constructed from dynamic random information such as MD can capture phonon information. Harmonic phonon calculations were conducted using both DFT and ACE potentials as the engines for interatomic force calculations within ALAMODE. Figure 2 shows the characteristics of harmonic phonons, such as the dispersion and density of states, obtained using both methods.

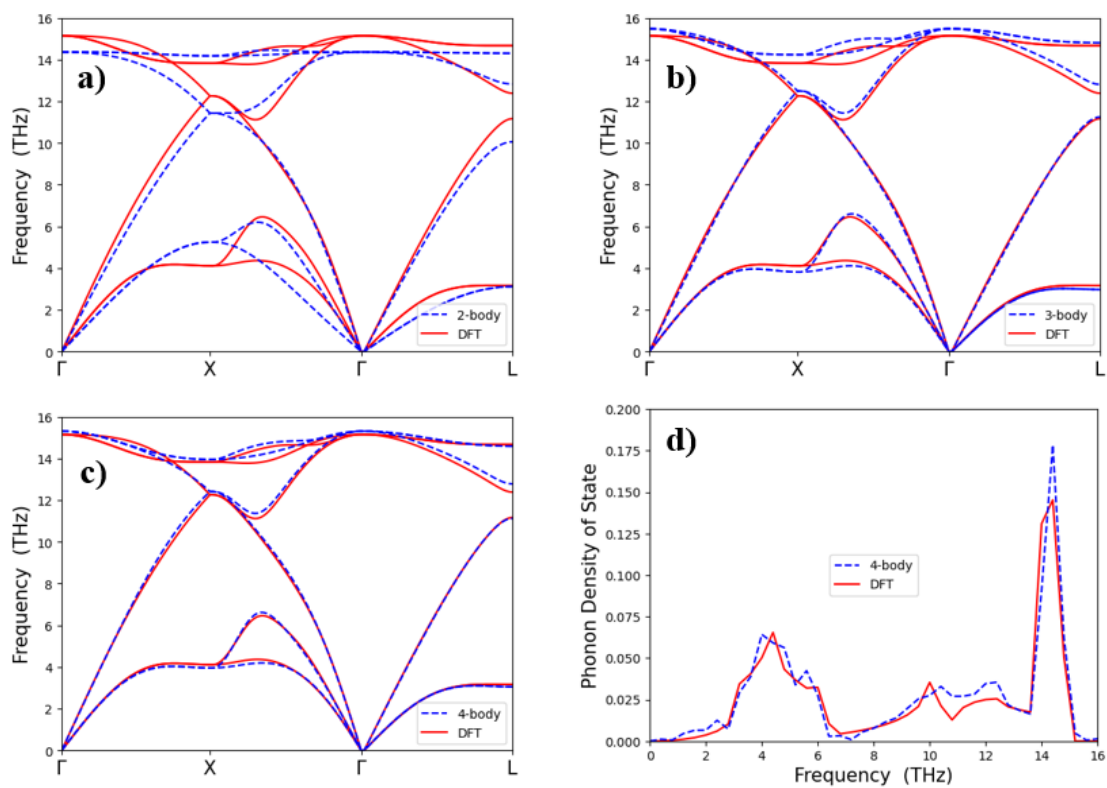

FIG. 2. Phonon dispersion and density of state in comparison with DFT results: Dispersion obtained by (a) 2-body, (b) 3-body, and (c) 4-body ACE potentials. (d) Vibration spectrum obtained by MD with a 4-body potential (blue dashed line). The phonon density of state obtained by DFT is indicated by red solid lines.

The phonon dispersion calculated with a 2-body ACE potential approximates that obtained using DFT. However, significant differences can be observed in the curve shapes, particularly in the acoustic phonons at frequencies below 12 THz (Figure 2a). At point X, the phonon dispersion obtained using the ACE potential exhibits a steep gradient, with a larger phonon velocity than the DFT, which tends to lower the accuracy of thermal conductivity calculations. In general, acoustic phonons considerably affect the thermal conductivity in Si [40], and thus, the reproducibility of acoustic phonons was carefully considered (Figure S3-1 in the Supporting Information). Nevertheless, the 3- and 4-body ACE potentials accurately reproduce the DFT dispersions (Figures 2b and 2c). Although the 3-body potential approximates the shape of the acoustic phonon dispersion, a slight shift can be observed at point X. The 4-body potential further enhances the accuracy at this point, thereby improving the reproducibility. In the calculation of the average energy of the acoustic phonon, increasing the expansion order allows the ACE potential to approach the DFT results, with the 4-body potential value at 4.35 THz matching the DFT value at 4.37 THz (Table S5-1). Additionally, the accuracy of optical phonon reproduction is enhanced as the expansion order increases, albeit with larger discrepancies compared with acoustic phonons. Higher-order expansion coefficients may be required to reproduce the opposite phase oscillations of optical phonons. However, our analysis mainly focused on acoustic phonons. The 3- and 4-body potentials can accurately reproduce the dispersions in harmonic phonon calculations, in which instantaneous forces are calculated. Next, we investigated whether the oscillation characteristics were maintained in dynamic MD calculations. Figure 2(d) presents a comparison of the vibration spectra obtained by bulk Si MD with a 4-body potential (Equation 1) and the PDOS of DFT. The acoustic phonon peaks below 12 THz are well replicated, indicating strong agreement across the entire q-space and not only along symmetry axes such as Γ-X-Γ-L (refer to Figure 5S-1 in the Supporting Information for the 2-body and 3-body potential cases). Therefore, the 3-body and 4-body Si ACE potentials contain highly accurate vibrational information.

  The next stage involved the validation of the thermal conductivity derived from anharmonic phonon calculations. In such calculations, ALAMODE accounts for the system symmetry and computes interatomic forces using 20 patterns of diatomic displacements ($d = 0.04$ Å). In general, the use of ACE potentials as the force calculation engine allows anharmonic phonon calculations to be performed 6000 times faster than the DFT approach. Anharmonic phonon calculations enable the computation of phonon lifetimes, which can then be used to calculate thermal conductivity using the Boltzmann equation [46–48]. Figure 3a shows the thermal conductivity results calculated using DFT and ACE potentials with different expansion orders.

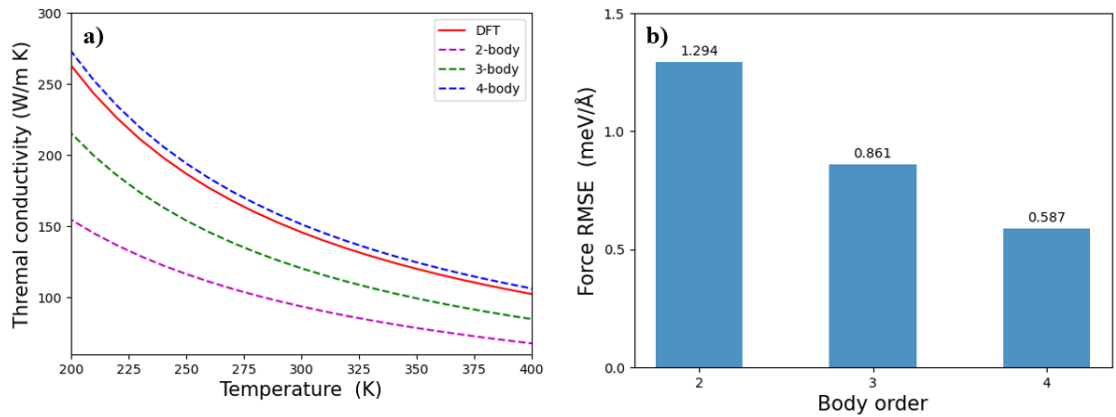

FIG. 3. (a) Si thermal conductivity calculated using the Boltzmann equation in the DFT framework and 2-body, 3-body, and 4-body ACE potentials. (b) RMSEs of forces in all directions ($F_x$, $F_y$, and $F_z$) obtained by DFT and ACE potentials in anharmonic calculations.

The thermal conductivity of the 2-body ACE potential is highly deviated from the DFT results. This discrepancy is attributable to the deviation in phonon dispersions (Figure 2a), which leads to differences in phonon velocities and energy. The thermal conductivities of the 3-body and 4-body potentials, which have similar phonon dispersions (Figures 2b and 2c), differ in accuracy from the DFT results. The thermal conductivity obtained using the 4-body potential is consistent with the DFT results, whereas that obtained using the 3-body potential is significantly different. This observation indicates that an expansion order higher than 4 must be employed when calculating the interatomic forces of anharmonic phonons [50]. The ACE potentials incorporate physical representations of higher-order interatomic force constants. Thus, the RMSEs for the interatomic forces in anharmonic phonon calculations were analyzed to clarify the difference in accuracy between the 3-body and 4-body potentials. Figure 3b shows the RMSE of forces in the $x$, $y$, and $z$ directions (also refer to Figure S6-1). The RMSE of the 2-body is 1.294 meV/Å, indicating low accuracy. The RMSE of the 3-body and 4-body ACE potentials are 0.861 and 0.587 meV/Å, respectively, with a difference of 0.273 meV/Å. This difference likely contributes to the discrepancy in the thermal conductivity accuracy relative to that obtained using DFT. These outcomes demonstrate that an accuracy of the order of 0.1 meV/Å is required for the interatomic forces when simulating thermal conductivity using an MLFF.

The calculation results were also compared with the experimentally obtained Si single-crystal thermal conductivity at 300 K. The experimental values range between 148 and 156 W/m K [9,38–40], whereas the DFT value in the current study is 145 W/m K and that from the 4-body ACE potential is 151 W/m K. The MLFF results appear to align more closely with the experimental values. However, because the MLFF relies on DFT results as training data, it cannot be more accurate than DFT. It is only in this case that the deviation from the DFT results happens to be closer to the experimental values.

The analytical results demonstrates that the 4-body ACE potential constructed via AIMD can reproduce vibrational behavior with high DFT-level accuracy. To validate the accuracy and calculation speed of the ACE potential, we applied it to a large-scale calculation involving over 100,000 atoms, as described in the subsequent section. This demonstration was aimed at highlighted the applicability of ACE potentials to thin films and vacancy effect calculations in NEMD.

### C. Application of ACE potentials to NEMD simulations

This section describes the application of the proposed ACE potentials. As discussed, ACE potentials are characterized by a high accuracy and calculation speed and can thus likely yield accurate results in large-scale computations. We selected the NEMD calculation of the thermal conductivity in thin films as an illustrative example. Details regarding the models and calculation conditions for NEMD are presented in Section S2 of the Supporting Information and Section II. We observed the variations in the thermal conductivity with the film thickness $L$ of the thermal transport region of Si ranging from 100 nm to 500 nm using NEMD and examined whether the results were consistent with those of DFT calculations at 300 K. For comparison with DFT results, we used the following approximation equation corresponding to the accumulative lattice thermal conductivity [46]:

$$\kappa(L) = \frac{1}{VN} \sum_{q,j} c_{qj} v_{qj} v_{qj} \tau_{qj} \Theta(L - |v_{qj}|\tau_{qj}), \quad (3)$$

where $c_{qj}$, $v_{qj}$, and $\tau_{qj}$ represent the specific heat, group velocity, and lifetime of wavenumber $q$ and phonon mode $j$, respectively. The term $\Theta(L-|v_{qj}|\tau_{qj})$ is a step function, which indicates that only the phonon modes whose mean free path (MFP) is smaller than the film thickness $L$ contribute to the total thermal conductivity. The MFP of phonons in the Si crystal is illustrated in Figure S7-1 in the Supporting Information. The phonons with MFPs between 100 nm and 500 nm are concentrated in the range of 2–7 THz (Figure S7-1), which indicates that the dependence of the thermal conductivity on the film thickness is likely influenced by vibrations within this range. Indeed, as the film thickness varies from $L$ = 100 nm to $L$ = 500 nm, the cumulative thermal conductivity by DFT ranges from 39.1 W/m K to 71.4 W/m K. These values were compared with the thermal conductivity obtained from NEMD.

Figure 4a shows the thermal conductivity obtained using NEMD with ACE potentials of different expansion orders and different $L$ values.

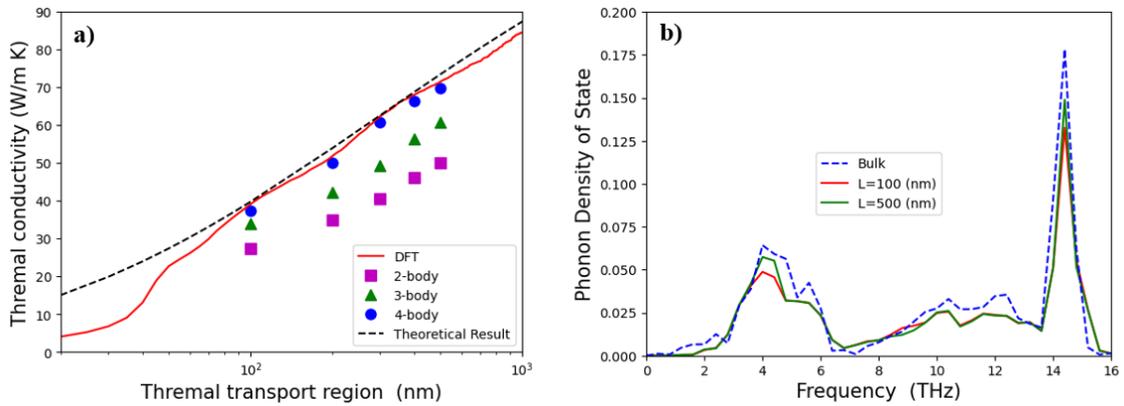

FIG. 4. (a) Variation in the Si thermal conductivity with the thermal transport region length at 300 K. The approximate DFT values obtained using Equation 3 are compared with the NEMD obtained values using 2-, 3-, and 4-body potentials. (b) Vibration spectra comparing bulk and finite length systems, $L$ = 100 nm and 500 nm, with a 4-body ACE potential.

The thermal conductivity determined by the 2- and 3-body ACE potentials monotonically increases with the film thickness, similar to the DFT results. However, the thermal conductivity ranges from 27.2 to 50.1 W/m K for the 2-body potential and from 33.8 to 60.8 W/m K for the 3-body potential as the thickness varies from 100 nm to 500 nm. These values are smaller than those obtained using the DFT, which highlights the necessity of improving the potential in NEMD to an accuracy of 0.1 meV/Å, as in the case of the anharmonic phonon calculations described in Section III B. The thermal conductivity values obtained using the 4-body potential range from 37.4 to 69.5 W/m K, consistent with the DFT results. This finding suggests that the 4-body ACE potential is sufficient for assessing the film thickness dependence of the thermal conductivity of Si. Next, we analyzed the vibration spectra to observe changes in the vibrational behavior with $L$ when using the 4-body ACE potential (Figure 4b). The DFT phonon MFPs shown in Figure S7-1 indicate that phonons with MFPs ranging from 100 nm to 500 nm are concentrated in the frequency range of 2–7 THz. We compared the changes in the vibrational spectra in this range to those of the bulk and the system with $L$=100 nm and $L$=500 nm, obtained using the 4-body potential. The vibration spectrum around 4 THz decreases when changing the finite thickness is altered, indicating reduced atomic vibrations with wavelengths in this range. As $L$ increases to 500 nm, the 4 THz peak approximates that in the bulk spectrum. Overall, the ACE potential can accurately calculate the thermal conductivity compared with DFT and large-scale simulations, while preserving the physical vibrational behavior.

Furthermore, we compared our results with those of the Callaway–Holland theoretical model for thermal conductivity, which is commonly used in the experimental analyses of Si thin films [39,40,43–45]. Details of this model can be found in Section S3 in the Supporting Information. The dependence of the thermal conductivity on the film thickness, as obtained from the 4-body ACE potential, is consistent with the results of the theoretical equation and DFT. This finding confirms that the ACE potential, even one constructed using a small AIMD dataset (3000 data points), can reliably simulate the thermal conductivity.

Lastly, as an exploratory application, we performed NEMD calculations with defects, using the 4-body ACE potential on the $L$ = 100 nm model. In general, standard NEMD calculations with ACE potentials built from bulk MD data are not suitable in scenarios involving defect states, as the interatomic forces can rapidly increase during the MD step, leading to calculation breakdown. To address this problem, we incorporated additional MD data involving a silicon defect vacancy ($V_{Si}$) into the ACE potential. The additional training data consisted of MD data with a single defect vacancy introduced into a 64-atom bulk dataset (resulting in 63 atoms) at temperatures of 100 K, 300 K, and 500 K. The cutoff radius for constructing the ACE potential was set as 5 Å. Figure 5a illustrates the computational model of NEMD for this training data environment. We introduced random defects into the thermal transport region while ensuring that the distance between defect vacancies is not less than $2R$, considering periodic boundary conditions (Figure 5b).

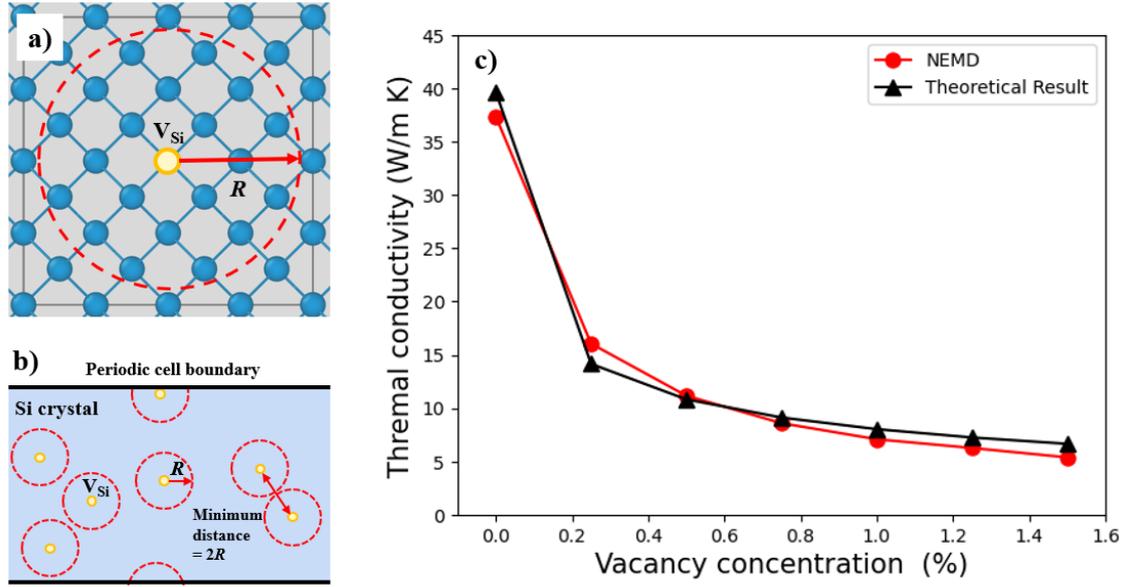

FIG. 5. (a) Schematic of the training data with Si defect $V_{Si}$ and a cell length of 10.8 Å. The radius around $V_{Si}$, $R$ = 5 Å, corresponds to the cutoff radius for neighboring constructed potentials. (b) Schematic of NEMD model for defect creation in the thermal transport region. (c) Thermal conductivity as a function of defect rate in the $L$ = 100 nm model.

The model with defect vacancies was used to examine the effect of introducing defects at rates of up to 1.5% on the thermal conductivity. The results are shown in Figure 5c. The introduction of defects decreases the thermal conductivity, consistent with previously reported findings [11,15,23]. For example, the thermal conductivity is 15.2 W/m K for a defect rate of 0.2%, which is approximately 60% lower than that of a perfect crystal. This reduction can be attributed to the substantial alteration in phonon states, such as the lifetime, caused by scattering at the defect sites. Further details regarding the theoretical model can be found in Section S3 in the Supporting Information. The results for defect rate dependence are consistent with those of theoretical equations such as the Callaway–Holland model. The deviation between the theoretically obtained and NEMD computed results is within 5%. The theoretical calculations indicate that defects affect phonon lifetimes, resulting in decreased thermal conductivity. NEMD computations with ACE potentials can successfully reproduce this reduction in the phonon lifetime. These results highlight that using simple training data, ACE potentials can be extended to realize large, complex calculations with an accuracy comparable to that of DFT. Notably, comparison with experimental data requires the incorporation of detailed training data pertaining to defect states [49], which is difficult to achieve with this simplified defect model. Nevertheless, no discrepancy in order or trend as a decrease in the thermal conductivity of the model with the introduction of defects is observed [11,15,23].

The illustrative examples of film thickness and defect dependencies highlight that calculations based on ACE potentials do not require complex theoretical models and can be accomplished with high precision for systems containing over 100,000 atoms. We believe that MLFF MD approaches are better suited for guiding device design than theoretical models because they can incorporate defect structures and other information, such as the density, local compositional fluctuations, and grain effect, thereby yielding accurate results more closely aligned with experimental data.

## IV. CONCLUSIONS

This study demonstrated the capability of NEMD in simulating the thermal conductivity in large systems consisting of more than 100,000 atoms and achieving DFT-level accuracy using an MLFF. The ACE force field was used as the MLFF, as it can simultaneously provide high levels of computational accuracy and speed. Unlike neural network based MLFFs, ACE incorporates physical representations, which can enable the derivation of physical insights.

An ACE force field for Si was constructed and validated through comparison with existing frameworks. The training data included random structural and interatomic force data obtained using first-principles MD. The thermal conductivity was calculated with ACE potentials of a 4-body expansion order, using phonons for accuracy verification and NEMD simulations for large-scale computational verification. The accuracy of the proposed ACE potentials was verified, and the results demonstrated that to accurately calculate the thermal conductivity in phonon calculations, the interatomic force accuracy must be of the order of 0.1 meV/Å with an atomic displacement of 0.01 Å. The 4-body ACE potential results satisfied this accuracy level and were consistent with the DFT findings. NEMD calculations for systems with more than 100,000 atoms were performed to illustrate the application of the constructed force field to a large scale. The considered systems included a thin film model with a thickness ranging from 100 nm to 500 nm and a 100 nm thin film model with defects. The 4-body ACE potential successfully reproduced the DFT results with high accuracy and exhibited a reasonable agreement with the theoretically calculated values. Thus, we concluded that ACE offers both high accuracy and speed in practical scenarios, such as thermal conductivity calculations.

The accuracy of force field computations has historically been a challenge in MD calculations. However, the use of the ACE potential avoids the necessity of labor-intensive experiments and complex parameter fitting methods, and it can be constructed in a straightforward manner. Instead of deriving parameters from experimental data, they can be fitted from data calculated through NEMD simulations. The ACE-NEMD approach can facilitate the formulation of design guidelines for device development, as it can incorporate atomic-level features, thereby ensuring that the simulation results match the experimental observations. Given the DFT-level accuracy and high speed of the ACE potential, future research can be aimed at extending its application to other scenarios. Furthermore, ACE simulations can be scaled with parallel CPUs to perform large-scale computations for various scenarios, such as the analysis of interface roughness structures, interface thermal conduction, and effects of defects and grains. By using high-performance computing machines, the relevant information can be integrated with micrometer-scale simulations, consistent with the sizes of actual devices.

# REFERENCES

[1] Q. Zheng, M. Hao, R. Miao, J. Schaadt, and C. Dames, Advances in thermal conductivity for energy application: a review. Prog. Energy. **3**, 012002 (2021)

[2] W. Liu, Q. Jie, H. S. Kim, and Z. Ren, Current progress and future challenges in thermos electric power generation: from materials to devices. Acta Mater. **87**, 357–376 (2015).

[3] K. Biswas, J. He, I. D. Blum, C.-I. Wu, T. P. Hogan, D. N. Seidman, V. P. Dravid, and M. G. Kanatzidis, High-performance bulk thermoelectrics with all-scale hierarchical architectures. Nature. **489**, 414–418 (2012).

[4] D. Luo, Z. Liu, Y. Yan, Y. Li, R. Wang, L. Zhang, and X. Yang, Recent advances in modeling and simulation of thermoelectric power generation. Energy Convers. Manag. **273**, 116389 (2022)

[5] X. Gu, Z. Fan, and H. Bao, Thermal conductivity prediction by atomistic simulation method: Recent advances and detailed comparison. J. Appl. Phys. **130**, 210902 (2021).

[6] X. Chen, D. Parker, and D. J. Singh, Acoustic impedance and interface phonon scattering in $Bi_2Te_3$ and other semiconducting materials. Phys. Rev. B. **87**, 045317 (2013).

[7] G. Ding, G. Gao, and K. Yao, High-efficient thermoelectric materials: The case of orthorhombic IV-VI compounds. Sci. Rep. **5**, 9567 (2015).

[8] H. Wang, Y. Gao, and G. Liu, Anisotropic phonon transport and lattice thermal conductivities in tin dichalcogenides $SnS_2$ and $SnSe_2$. RSC Adv. **7**, 8098–8105 (2017)

[9] E. Chávez-Ángel et al., Reduction of the thermal conductivity in free-standing silicon nano-membranes investigated by non-invasive Raman thermometry. APL Mater. **2,** 012113 (2014).

[10] Y. S. Ju and K. E. Goodson, Phonon scattering in silicon films with thickness of order 100nm. Appl. Phys. Lett. **74**, 3005–3007 (1999).

[11] Y. Lee, S. Lee, and G. S. Hwang, Effects of vacancy defects on thermal conductivity in crystalline silicon: A nonequilibrium molecular dynamics study. Phys. Rev. B **83**, 125202 (2011).

[12] Z. Wang, J. E. Alaniz, W. Jang, J. E. Garay, and C. Dames, Thermal conductivity of nanocrystalline silicon: Importance of grain size and frequency-dependent mean free paths. Nano. Lett. **11**, 2206–2213 (2011).

[13] W. Zhu, G. Zheng, S. Cao, and H. He, Thermal conductivity of amorphous $SiO_2$ thin film: A molecular dynamics study. Sci. Rep. **8**, 10537 (2018).

[14] D. P. Sellan, E. S. Landry, J. E. Turney, A. J. H. McGaughey, and C. H. Amon, Size effects in molecular dynamics thermal conductivity predictions. Phys. Rev. B **81**, 214305 (2010).

[15] Z. Xingli and S. Zhaowei, Effects of vacancy structural defects on the thermal conductivity of

silicon thin films. J. Semicond. **32**, 053002 (2011).

[16] Y. Lee, S. Lee and G. S. Hwang, Effects of vacancy defects on thermal conductivity in crystalline silicon: A nonequilibrium molecular dynamics study. Phys. Rev. B **83**, 125202 (2011).

[17] H. Zaoui, P. L. Palla, F. Cleri, and E. Lampin. Length dependence of thermal conductivity by approach-to-equilibrium molecular dynamics. Phys. Rev. B **94**, 054304 (2016).

[18] J. S. Alexander, C. Maxwell, J. Pencer, and M. Saoudi, Equilibrium molecular dynamics calculations of thermal conductivity: a "how-to" for the beginners. CNL Nucl. Rev. **9**, 1, (2020).

[19] S. Fujii and A. Seko, Structure and lattice thermal conductivity of grain boundaries in silicon by using machine learning potential and molecular dynamics. Comp. Mat. Sci. **204**, 111137 (2022).

[20] X. Qian, S. Peng, X. Li, Y. Wei, and R. Yang, Thermal conductivity modeling using machine learning potentials: application to crystalline and amorphous silicon. Mater. Today Phys. **10**, 100140 (2019).

[21] C. Verdi, F. Karsai, P. Liu, R. Jinnouchi, and G. Kresse, Thermal transport and phase transitions of zirconia by on-the-fly machine-learned interatomic potentials. NPJ Comput. Mater. **7**, 156 (2021).

[22] B. Mortazavi, E. V. Podryabinkin, I. S. Novikov, S. Roche, T. Rabczuk, X. Zhuang, and A. V. Shapeev, Efficient machine-learning based interatomic potentials for exploring thermal conductivity in two-dimensional materials. J. Phys. Mater. **3**, 02LT02 (2020)

[23] H. Babaei, R. Guo, A. Hashemi, and S. Lee, Machine-learning-based inter atomic potential for phonon transport in perfect crystalline Si and crystalline Si with vacancies. Phys. Rev. Mater. **3**, 074603 (2019).

[24] S. Batzner, A. Musaelian, L. Sun, M. Geiger, J. P. Malioa, M. Kornbluth, N. Molinari, T. E. Smidt, and B. Kozinsky, E(3)-equivariant graph neural networks for data-efficient and accurate interatomic potentials. Nat. Commun. **13**, 2453 (2022).

[25] S. Takamoto *et al.*, Towards universal neural network potential for material discovery applicable to arbitrary combination of 45 elements. Nat. Commun. **13**, 2991 (2022).

[26] A. P. Bartók and G. Csányi, Gaussian approximation potentials: A brief tutorial introduction. Int J. Quantum Chem. **115**, 1051–1057 (2015).

[27] A. P. Bartók, M. C. Payne, R. Kondor and G. Csányi, Gaussian approximation potentials: The accuracy of quantum mechanics, without the electrons. Phys. Rev. Lett. **104**, 136403 (2010).

[28] A. V. Shapeev, Moment potentials: a class of systematically improvable interatomic potentials. Multiscale Model. Simul. **14**, 1153–1173 (2016).

[29] I. S. Novikov, K. Gubaev, E. V. Podryabinkin, and A. V. Shapeev, The MLIP package: Moment tensor potentials with MPI and active learning. Sci. Technol. **2**, 025002 (2021).

[30] R. Drautz, Atomic cluster expansion for accurate and transferable interatomic potentials. Phys. Rev. B **99**, 014104 (2019).

[31] Y. Lysogorskiy *et al.*, Performant implementation of the atomic cluster expansion (PACE) and

application to copper and silicon. NPJ Comput. Mater. **7**, 97 (2021).

[32] A. Bochkarev, Y. Lysogorskiy, S. Menon, M. Qamar, M. Mrovec, and R. Drautz, Efficient parametrization of the atomic cluster expansion. Phys. Rev. Mater. **6**, 013804 (2022).

[33] G. Dusson, M. Bachmayr, G. Csányi, R. Drautz, S. Etter, C. van der Oord, and C. Ortner, Atomic cluster expansion: Completeness, efficiency and stability. J. Comput. Phys. **454**, 110946 (2022).

[34] K. Shimamura, Y. Takeshita, S. Fukushima, A. Koura, and F. Shimojo, Estimating thermal conductivity of a-Ag2Se using ANN potential with Chebyshev descriptor. Chem. Phys. Lett. **778**, 138748 (2021).

[35] P. Giannnozzi *et al.*, Quantum Espresso: A modular and open-source software project for quantum simulations of materials. J.Phys. Condens.Matter **21**, 395502 (2009)

[36] P. Giannozzi, *et al.*, Advanced capabilities for materials modelling with Quantum Espresso. J.Phys. Condens.Matter **29**, 465901 (2017).

[37] S. Plimpton, Fast parallel algorithms for short-range molecular dynamics. J. Comput. Phys. **117**, 1 (1995).

[38] M. Asheghi, Y. K. Leung, S. S. Wong, and K. E. Goodson, Phonon-boundary scattering in thin silicon layers. Appl. Phys. Lett. **71**, 1798 (1997).

[39] J. Ma, W. Li, and X. Luo. Examining the Callaway model for lattice thermal conductivity. Phys. Rev. B. **90**, 035203 (2014).

[40] B. L. Davis and M. I. Hussein, Thermal characterization of nanoscale phononic crystals using supercell lattice dynamics. AIP Adv. **1**, 041701 (2011).

[41] R. Meyer and D. Comtesse, Vibrational density of states of silicon nanoparticles. Phys. Rev. B. **83**, 014301 (2011).

[42] P. J. Berryman, D. A. Faux, and D. J. Dunstan, Solvation pressure in ethanol by molecular dynamics simulations. Phys. Rev. B. **76**, 104303 (2007).

[43] J. Callaway, Model for lattice thermal conductivity at low temperatures. Phys. Rev. **114**, 1046–1051 (1959).

[44] A. M. Toxen, Lattice thermal conductivity of germanium-silicon alloy single crystals at low temperatures. Phys. Rev. **122**, 450–458 (1961).

[45] A. K. Suliman and M. S. Omar, Modified callaway model calculations for lattice thermal conductivity of a 20 nm diameter silicon nanowire. Exp. Theor. Nanotechnol. **5**, 65–76 (2021).

[46] T. Tadano, Y. Gohda, and S. Tsuneyuki, Anharmonic force constants extracted from first-principles molecular dynamics: applications to heat transfer simulations. J. Phys. Condens. Matter. **26**, 225402 (2014).

[47] P. E. Blöchl, O. Jepsen, and O. K. Andersen, Improved tetrahedron method for Brillouin-zone integrations. Phys. Rev. B. **49**, 16223–16233 (1994).

[48] K. Esfarjani, G. Chen, and H. T. Stokes, Heat transport in silicon from first-principles calculations.

Phys. Rev. B. **84**, 085204 (2011).

[49] A. P. Bartók, J. Kermode, N. Bernstein, and G. Csányi, Machine learning a general-purpose interatomic potential for silicon. Phys. Rev. X. **8**, 041048 (2018).

[50] D. A. Broido, M. Malomy, G. Birner, N. Mingo, and D. A. Stewart, Intrinsic lattice thermal conductivity of semiconductors from first principles. Appl. Phys. Lett. **91**, 231922 (2007).

[51] E, Pop, R. W. Duttom, and K. E. Goodson, Analytic band Monte Carlo model for electron transport in Si including acoustic and optical phonon dispersion. J. Appl. Phys. **96**, 4998 (2004)

**ACKNOWLEDGMENTS**

This study received no funding.

# Supporting information for

# Atomic cluster expansion force field based thermal property material design with density functional theory level accuracy in non-equilibrium molecular dynamics calculations over sub-million atoms

Takumi Araki[1*], Shinnosuke Hattori[1], Toshio Nishi[1], and Yoshihiro Kudo[1]

[1]Material Device Analysis Center, Third Research Department, Sony Semiconductor Solutions Corporation

*Corresponding author: Takumi.Araki@sony.com

## S1. Parameters for the Si ACE potential development.

TABLE S1-1. Pairs of maximum quantum numbers $n_{max}$ and $l_{max}$ for different cluster expansion orders $v$ used in the construction of the ACE potential for Si and number of parameters after construction. Details can be found in Ref [30].

| Expansion order v | rc [Å] | $n_{max}$ | $l_{max}$ | Number of parameters |
|---|---|---|---|---|
| 2 | 5.0 | 15 | 0 | 30 |
| 3 | 5.0 | 15/8 | 0/3 | 798 |
| 4 | 5.0 | 15/8/6 | 0/3/3 | 2414 |
| 5 | 5.0 | 15/8/6/3 | 0/3/3/2 | 3314 |
| 6 | 5.0 | 15/8/6/3/2 | 0/3/3/2/2 | 3698 |
| 7 | 5.0 | 15/8/6/3/2/1 | 0/3/3/2/2/1 | 3706 |

## S2. Si NEMD model.

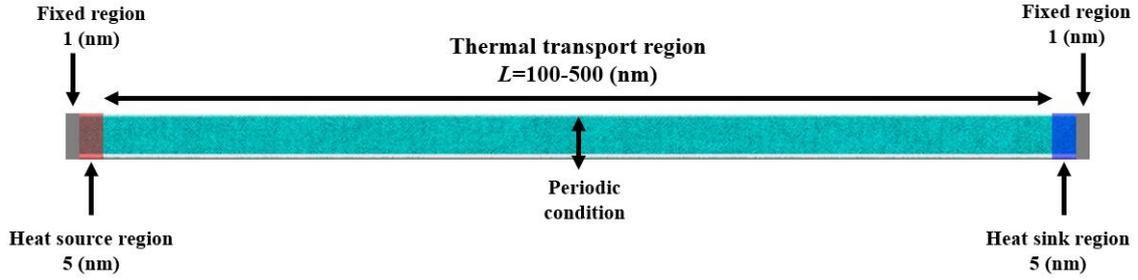

FIG. S2-1. Schematic of the Si NEMD model. The gray part indicates the fixed region. The red and blue parts represent the heat source and sink, respectively.

TABLE S2-1. MD calculation model for NEMD. Fixed regions were placed on both sides.

| Region | Length (nm) | Number of atoms |
|---|---|---|
| Fix×2 | 1 | 1920 |
| Heat source | 5 | 4736 |
| Heat sink | 5 | 4736 |
| Thermal transport | 100 | 94,208 |
| | 200 | 188,416 |
| | 300 | 282,624 |
| | 400 | 376,832 |
| | 500 | 471,040 |

## S3. Callaway–Holland model.

The Callaway–Holland model calculates the thermal conductivity $\kappa$ as [40]

$$\kappa = \frac{4\pi/3}{(2\pi)^3} \sum_{\mu} \int_0^{2\pi/a} C_{ph}(q,\mu) v^2(q,\mu) \tau(q,\mu) q^2 dq \quad , \quad (S1).$$

$$C_{ph}(q,\mu) = k_B \left(\frac{\hbar\omega(q,\mu)}{k_B T}\right)^2 \frac{\exp(\hbar\omega(q,\mu)/k_B T)}{[\exp(\hbar\omega(q,\mu)/k_B T) - 1]^2} \quad , \quad (S2).$$

$$\omega(q,\mu) = \omega_0(\mu) + a(\mu)q + b(\mu)q^2 \quad , \quad (S3).$$

$$v(q,\mu) = \frac{\partial \omega(q,\mu)}{\partial q} \quad (S4).$$

Here, $C_{ph}$, $v$, and $\tau$ denote the heat capacity, phonon velocity and phonon scattering time, respectively, as functions of the wavenumber $q$ and phonon mode $\mu$. A traditional approximation approach was used considering the phonon energy $\omega$ [51]. The parameters $\omega_0$, $a$, and $b$ are listed in Table S3-1.

The phonon scattering relaxation time $\tau$ can be decomposed into three components, each corresponding to a specific scattering mechanism: Umklapp $\tau_U$, impurity $\tau_I$, and boundary $\tau_B$. The inverse of these variables were summed via Matthiessen's rule to obtain the overall scattering time $\tau$.

$$\tau = (\tau_U^{-1} + \tau_I^{-1} + \tau_B^{-1})^{-1} = (AT\omega^2 \exp(-B/T) + D\omega^4 + v(q,\mu)/L)^{-1}, \quad (S5).$$

The following parameters were set to compare the thin film thermal conductivity obtained using the NEMD and theoretical calculations: $A = 2.1\mathrm{e}^{-19}$ sK$^{-1}$, $B = 180$ K and $D = 1.32\mathrm{e}^{-45}$ s$^3$ [40]. Using these parameters, we demonstrated the dependence of the thermal conductivity on the boundary length $L$ at 300 K, considering the contributions of acoustic and optical phonons, as shown in Figure S3-1.

TABLE S3-1. Quadratic phonon dispersion coefficients for longitudinal acoustic (LA), transverse acoustic (TA), longitudinal optical (LO), and transverse optical (TO) modes (Ref [51]).

| Phonon mode | $\omega_0$ (10$^{13}$ rad/s) | $a$ (10$^5$ cm/s) | $b$ (10$^{-3}$ cm$^2$/s) |
|---|---|---|---|
| LA | 0.00 | 9.01 | -2.00 |
| TA | 0.00 | 5.23 | -2.26 |
| LO | 9.88 | 0.00 | -1.60 |
| TO | 10.20 | -2.57 | 1.11 |

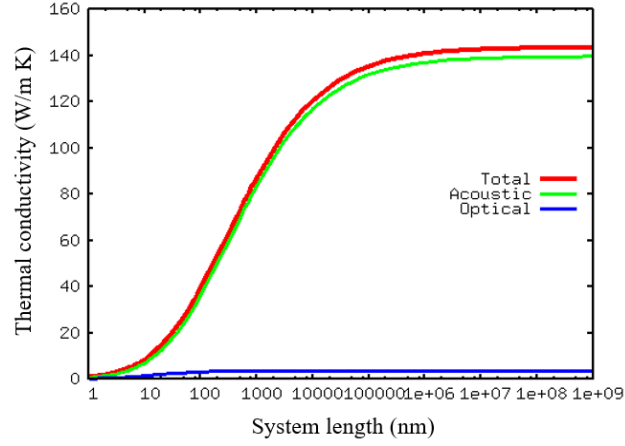

FIG. S3-1. Variations in the thermal conductivity of bulk Si with the boundary length $L$ in the $\Gamma$-X direction.

For comparison with the defect-introduced NEMD calculations, the formula for $\tau_I$ was changed to the following form [15]:

$$D = \frac{V_0}{4\pi v(q,\mu)^3} \sum_i f_i \left[1 - \left(\frac{M_i}{M}\right)\right]^2 , (S6).$$

where $V_0$ is the atomic volume, $M_i$ is the mass of an atom, $M$ is the average mass of all atoms, and $f_i$ is the fraction of atoms with mass $M_i$. For defect calculations, we set $M_i$ as zero and $f_i$ as the defect rate.

## S4. Computational resources required for performance evaluation with different numbers of atoms in terms of the simulation time per day.

We validated the computational performance of the Si ACE potential using the following setup. Sixty CPUs with the following configuration: Intel Xeon Gold 6248 20 cores 2.5 GHz were used to obtain computational performance data according to the number of atoms. A bulk configuration was adopted, with a time step of 0.1 fs, temperature of 300 K, and canonical NVT ensemble. Calculations were performed for 1000 MD steps, and the simulation time available per day was determined from the calculation time.

## S5. Properties of harmonic phonons in the ACE potential.

TABLE S5-1. Comparison of the average energy of acoustic and optical phonons calculated by DFT and 2-body, 3-body and 4-body Si ACE potential frameworks.

| Method | Acoustic phonon (THz) | Optical phonon (THz) |
|---|---|---|
| DFT | 4.37 | 14.31 |
| ACE 2-body | 4.15 | 14.02 |
| ACE 3-body | 4.30 | 14.61 |
| ACE 4-body | 4.35 | 14.45 |

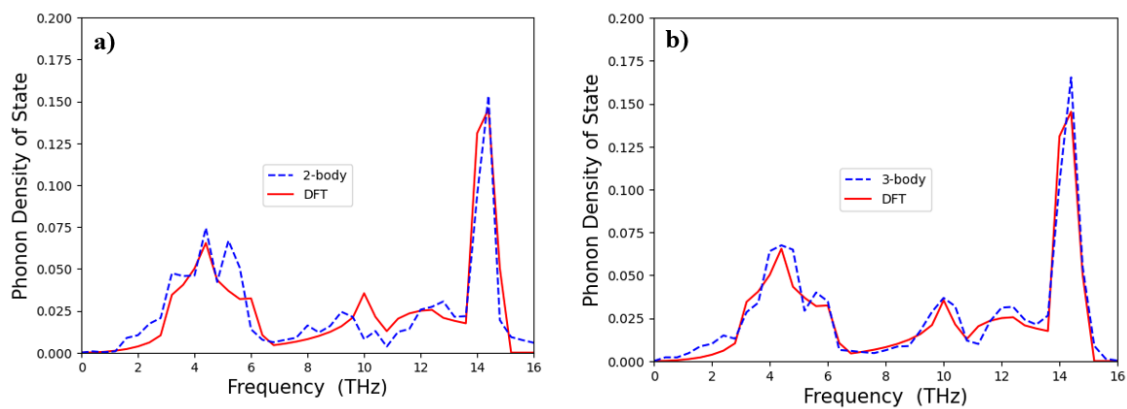

FIG. S5-1. Vibration spectrum obtained by MD with a) 2-body and b) 3-body potential (blue dashed line). The phonon density of state obtained by DFT is indicated by red solid lines.

## S6. Force RMSE for anharmonic calculations.

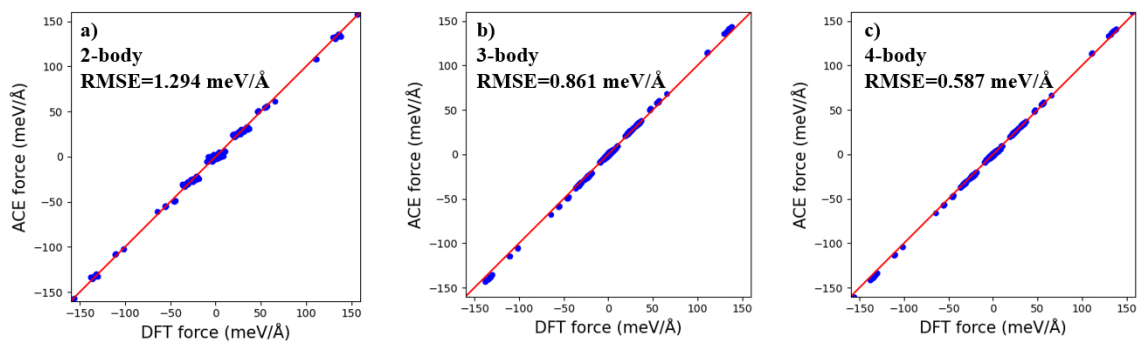

FIG. S6-1. Comparison of accuracy RMSE of interatomic forces derived from ACE potential and DFT in anharmonic phonon calculations for different expansion orders. a) 2-body, b) 3-body, and c) 4-body case.

## S7. Phonon mean free path in DFT calculation.

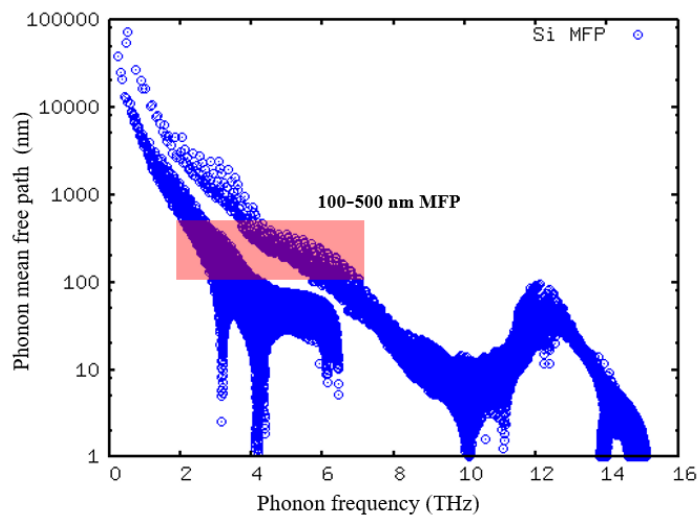

Figure 7S-1. Phonon mean free paths obtained from DFT calculations for a temperature of 300 K. The red area indicates phonons with mean free paths between 100 nm and 500 nm.